\documentclass{article} 
\usepackage{iclr2025_ai4na, times}

\usepackage{amsmath,amsfonts,bm}

\def\eqref#1{equation~\ref{#1}}

\def\1{\bm{1}}

\DeclareMathAlphabet{\mathsfit}{\encodingdefault}{\sfdefault}{m}{sl}
\SetMathAlphabet{\mathsfit}{bold}{\encodingdefault}{\sfdefault}{bx}{n}

\usepackage{hyperref}
\usepackage{url}
\usepackage{graphicx}
\usepackage{booktabs}
\usepackage{longtable}

\title{From Mutation to Degradation: Predicting Nonsense-Mediated Decay with NMDEP}

\author{Ali Saadat, Jacques Fellay \\
         School of Life Sciences\\
Ecole Polytechnique Fédérale de Lausanne\\
Lausanne, Switzerland\\
\texttt{\{ali.saadat, jacques.fellay\}@epfl.ch}}

\iclrfinalcopy 
\begin{document}

\maketitle

\begin{abstract}
Nonsense-mediated mRNA decay (NMD) is a critical post-transcriptional surveillance mechanism that degrades transcripts with premature termination codons, safeguarding transcriptome integrity and shaping disease phenotypes. However, accurately predicting NMD efficiency remains challenging, as existing models often rely on simplistic rule-based heuristics or limited feature sets, constraining their accuracy and generalizability. Using paired DNA and RNA data from The Cancer Genome Atlas, we benchmark embedding-only models and demonstrate that they underperform compared to a simple rule-based approach. To address this, we develop NMDEP (NMD Efficiency Predictor), an integrative framework that combines optimized rule-based methods, sequence embeddings, and curated biological features, achieving state-of-the-art predictive performance. Through explainable AI, we identify key NMD determinants, reaffirming established factors such as variant position while uncovering novel contributors like ribosome loading. Applied to over 2.9 million simulated stop-gain variants, NMDEP facilitates large-scale mRNA degradation assessments, advancing variant interpretation and disease research.

\end{abstract}

\section{Introduction}

\begin{figure}[h]
    \centering
    \includegraphics[width=0.95\textwidth]{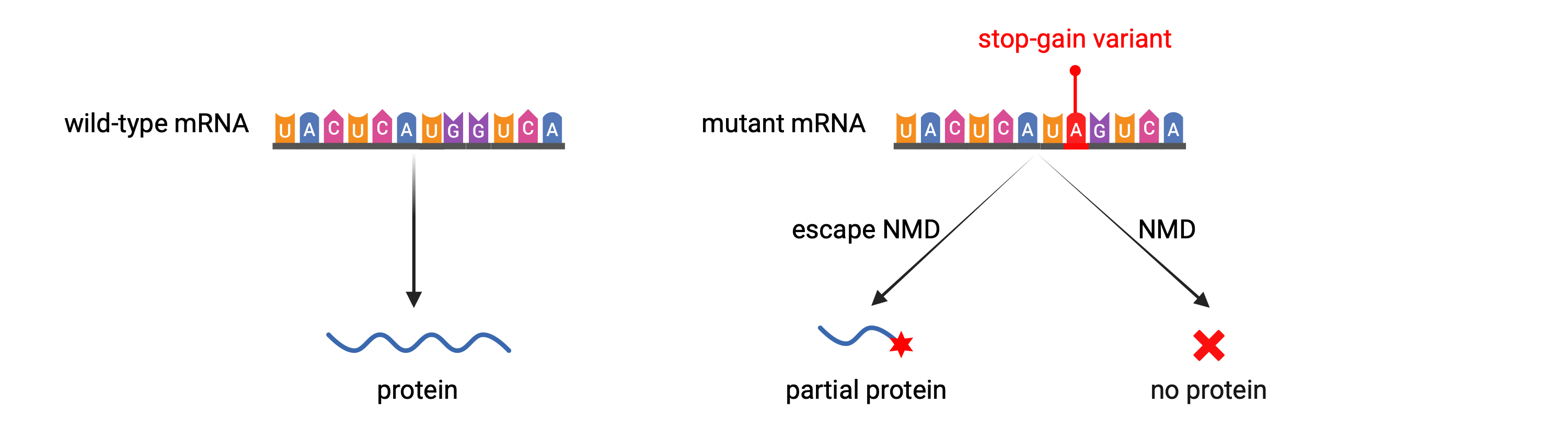} 
    \caption{\footnotesize Left: A wild-type mRNA is translated into a full-length protein. Right: An mRNA containing a stop-gain variant can either undergo NMD, resulting in no protein production, or escape NMD, leading to the translation of a truncated protein. Figure created with \url{BioRender.com}.}
    \label{NMD_process}
\end{figure}

Nonsense-mediated mRNA decay (NMD) is a crucial post-transcriptional regulatory mechanism that degrades mRNAs containing premature termination codons (PTCs) \citep{Chang2007-bi}. By selectively eliminating erroneous transcripts, NMD plays a vital role in maintaining transcriptome integrity, regulating gene expression, and influencing disease phenotypes \citep{Kurosaki2019-wt}. However, not all transcripts harboring PTCs undergo degradation, as some escape NMD due to context-dependent factors that modulate their stability (Figure \ref{NMD_process}). This variability complicates the computational prediction of NMD efficiency and its downstream effects on gene expression \citep{Supek2021-kw}.

Recent large-scale sequencing efforts, such as The Cancer Genome Atlas (TCGA) \citep{TCGA}, have enabled the systematic analysis of NMD efficiency across diverse genetic backgrounds \citep{Litchfield2020-mv}. In particular, allele-specific expression (ASE) patterns derived from paired DNA and RNA sequencing provide a means to quantify NMD efficiency at a single-variant resolution \citep{Rivas2015-ye}. 

Most existing methods have framed NMD efficiency prediction as a binary classification task (NMD vs. NMD escape), relying predominantly on rule-based heuristics \citep{Lindeboom2016-hb,Coban-Akdemir2018-nh,Lindeboom2019-kd,Klonowski2023-ax,Torene2024-ra} or limited feature sets \citep{Teran2021-nk,Kim2024-eh}, thereby restricting their accuracy and generalizability. Furthermore, these approaches often fail to account for context-dependent factors that modulate NMD efficiency. Consequently, there remains a critical need for more advanced models that incorporate rich sequence context and functional genomics data to enhance predictive performance.

To address these challenges, we explore the use of sequence embeddings from deep learning models trained on large-scale genomic data. We systematically evaluate multiple sequence aggregation strategies and compare embedding-based approaches with traditional feature-based models. Furthermore, we introduce NMDEP (NMD Efficiency Predictor), an integrative framework that combines optimized rule-based methods, deep-learning-derived embeddings, and curated biological features to achieve state-of-the-art predictive performance.

Our study benchmarks embedding-based methods for NMD efficiency prediction, showing that NMDEP outperforms rule-based and embedding-only models. Using explainable AI, we identify key NMD determinants, reaffirming known factors like variant position and uncovering novel contributors such as ribosome loading. Applying NMDEP to 2.9 million simulated stop-gain variants, we provide a large-scale resource for transcript stability assessments.

By integrating deep learning embeddings, curated biological features, and model interpretation, our framework advances NMD efficiency prediction, offering insights into transcriptome regulation and NMD escape mechanisms in disease.

\section{Methods}
\label{methods}

\subsection{Data collection}

\paragraph{NMD efficiency}

The NMD efficiency dataset was obtained from \citet{Kim2024-eh} which was constructed using paired DNA and RNA sequencing data from 9,235 samples in TCGA using 4,257 high-confidence nonsense variants. Since RNA-seq captures transcripts expressed in a cell, the variant allele frequency (VAF) observed in RNA sequencing (VAF$_{RNA}$) provides a measure of ASE, reflecting how much of a given allele's transcript is retained after cellular regulatory processes, including NMD. In contrast, the VAF of the same variant observed in DNA sequencing (VAF$_{DNA}$) represents the allele frequency prior to transcript-level regulation, making it a suitable baseline estimate of VAF$_{RNA}$ in the absence of NMD. To quantify NMD efficiency for each nonsense variant, the following equation was used: 
\( NMD_{\text{efficiency}} = -\log_2\left(\frac{\text{VAF}_\text{RNA}}{\text{VAF}_\text{DNA}}\right) \).

\paragraph{Data splitting}

We split the NMD efficiency dataset by chromosome to minimize the risk of information leakage. Chromosomes 20, 21, and 22 were assigned to the test set, chromosome 19 to the validation set, and the remaining chromosomes to the training set. This resulted in test, validation, and training splits of 5.5\%, 5.1\%, and 89.4\%, respectively.

\subsection{Benchmarking sequence embedding representations}

\paragraph{Study design}

\begin{figure}[h]
    \centering
    \includegraphics[width=0.95\textwidth]{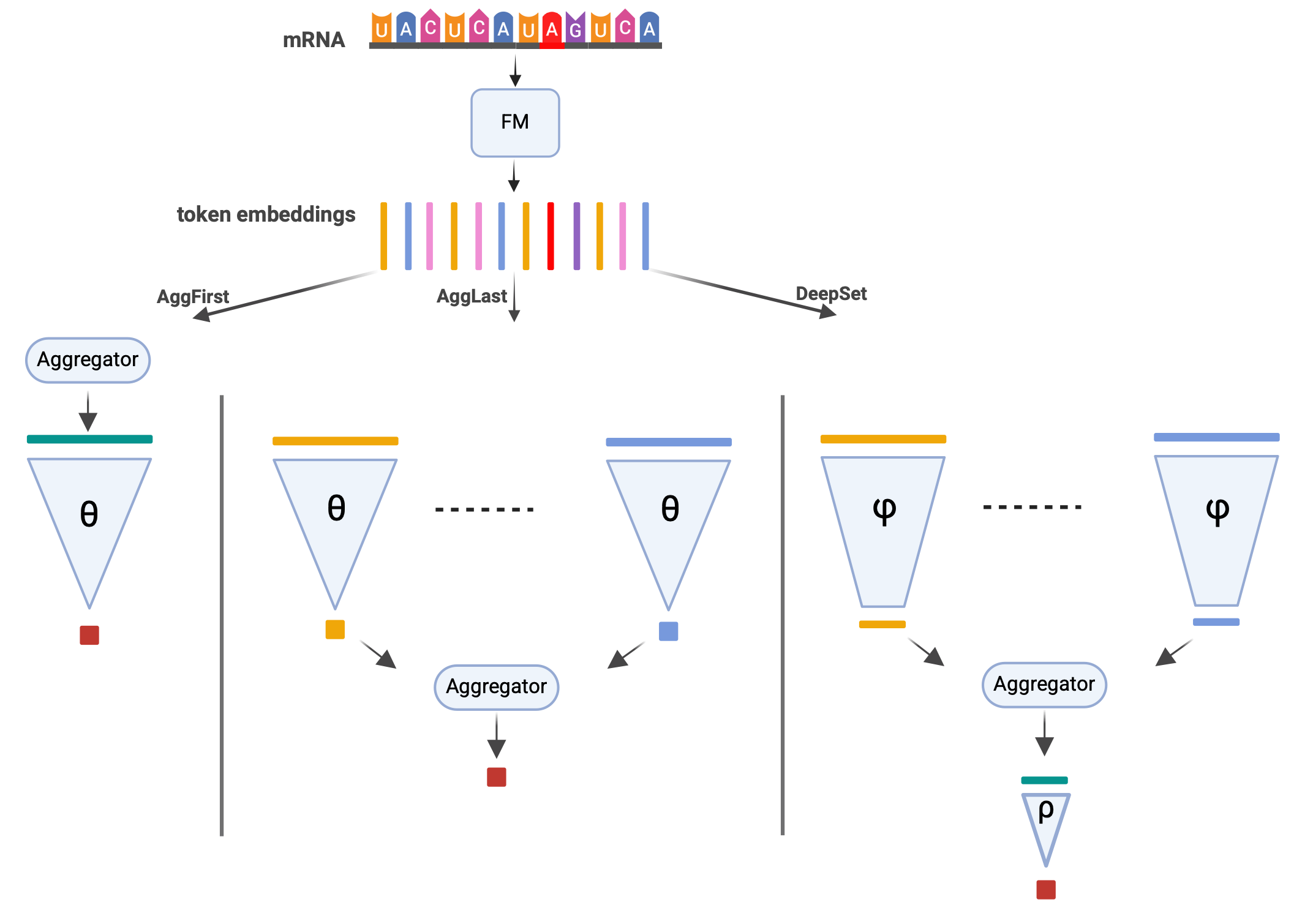} 
    \caption{\footnotesize Evaluation of three strategies for embedding aggregation: Aggregation First (AggFirst), Aggregation Last (AggLast), and DeepSet. Token embeddings were generated for reference (Ref) and alternative (Alt) mRNA sequences, using either Alt embeddings alone or the difference (Alt $-$ Ref). Four aggregation functions were tested: mean, max, sum, and token-level embedding. Figure created with \url{BioRender.com}}
    \label{NMD_orthrus}
\end{figure}

Since NMD efficiency prediction is a sequence-level phenotype influenced by a single nucleotide change, we conducted a systematic analysis to determine the optimal approach for predicting sequence-level outcomes from token embeddings. We evaluated three strategies: Aggregation First (AggFirst), Aggregation Last (AggLast), and a set neural network (DeepSet) \citep{Zaheer2017-mc}. Token embeddings were generated for both reference (Ref) and alternative (Alt) mRNA sequences, and we explored two encoding strategies: using only Alt token embeddings or computing the difference between Alt and Ref embeddings (Alt $-$ Ref). For each method, we tested four aggregation functions including mean, max, sum, and token (i.e., using the token embedding at the variant position). Figure \ref{NMD_orthrus} provides an overview of our benchmarking study design. 

For formulation, we define \( \mathbf{A} \) as the aggregation function. Given an mRNA sequence of length \( L \), each nucleotide is mapped to a token embedding in \( \mathbb{R}^K \). Thus, token embeddings are represented as: 
\( \mathbf{e}_i \in \mathbb{R}^K, \quad \forall i \in \{1, 2, \dots, L\} \), where \( \mathbf{e}_i \) denotes the token embedding of the \( i \)-th nucleotide.

\paragraph{AggFirst}

In this approach, the embeddings are first aggregated, then passed through the learnable module $\Theta$. In this approach, \(
NMD_{\text{efficiency}} = \Theta \left( \mathbf{A} \left( \{\mathbf{e}_i\}_{i=1}^{L} \right) \right)
\)

\paragraph{AggLast}

Here, we first use a shared, learnable \( \Theta \) to map each token embedding individually before aggregating them using \( \mathbf{A} \). Therefore, \(
NMD_{\text{efficiency}} = \mathbf{A} \left( \{\Theta(\mathbf{e}_i)\}_{i=1}^{L} \right)
\)

\paragraph{DeepSet}

In the third approach, each token embedding is first passed through a shared, learnable submodule \( \Phi \) to obtain a lower-dimensional representation. Next, the aggregation function \( \mathbf{A} \) is applied to combine the transformed embeddings into a single representation. Finally, another learnable submodule \( \rho \) is applied to the aggregated representation to predict NMD efficiency. This process can be formulated as: \(
NMD_{\text{efficiency}} = \rho \left( \mathbf{A} \left( \{\Phi(\mathbf{e}_i)\}_{i=1}^{L} \right) \right)
\)

\paragraph{Baseline}

As a baseline, we trained a model with similar architecture and hyper-parameters using four binary features based on previous studies \citep{Lindeboom2016-hb,Lindeboom2019-kd}:  
(1) \textbf{Last exon}: The stop-gain variant is located in the last exon of the transcript.  
(2) \textbf{Penultimate rule}: The nonsense variant occurs within the last 50 bp of the penultimate exon.  
(3) \textbf{Close to start}: The stop-gain variant is within 150 nucleotides of the start codon.  
(4) \textbf{Long exon}: The stop-gain variant is located in an exon longer than 407 bp.

\paragraph{Metrics}

During training, the model optimizes Mean Squared Error Loss (MSELoss). For evaluation, we compute five metrics: Mean Absolute Error (MAE), Root Mean Squared Error (RMSE), \( R^2 \), Spearman correlation, and Pearson correlation.

\paragraph{Hyper-parameters and software}

For mRNA sequence extraction and processing, we used \citet{genomekit}. Token embeddings were computed using Orthrus \citep{Fradkin2024-rr}, a pretrained, Mamba-based mRNA foundation model. Model training was conducted in PyTorch \citep{Paszke2019-va} with a learning rate of \(5 \times 10^{-4}\) and weight decay of \(5 \times 10^{-4}\). The functions \( \Theta \), \( \Phi \), and \( \rho \) were implemented as multi-layer perceptrons (MLPs) with two hidden layers. Specifically, the hidden layers of \( \Theta \) and \( \Phi \) had a size of 8, while \( \rho \) had a hidden dimension of 4. To mitigate overfitting, we applied a dropout rate of 0.25 and trained the models for up to 50 epochs with early stopping.

\subsection{NMDEP}

\begin{figure}[h]
    \centering
    \includegraphics[width=0.95\textwidth]{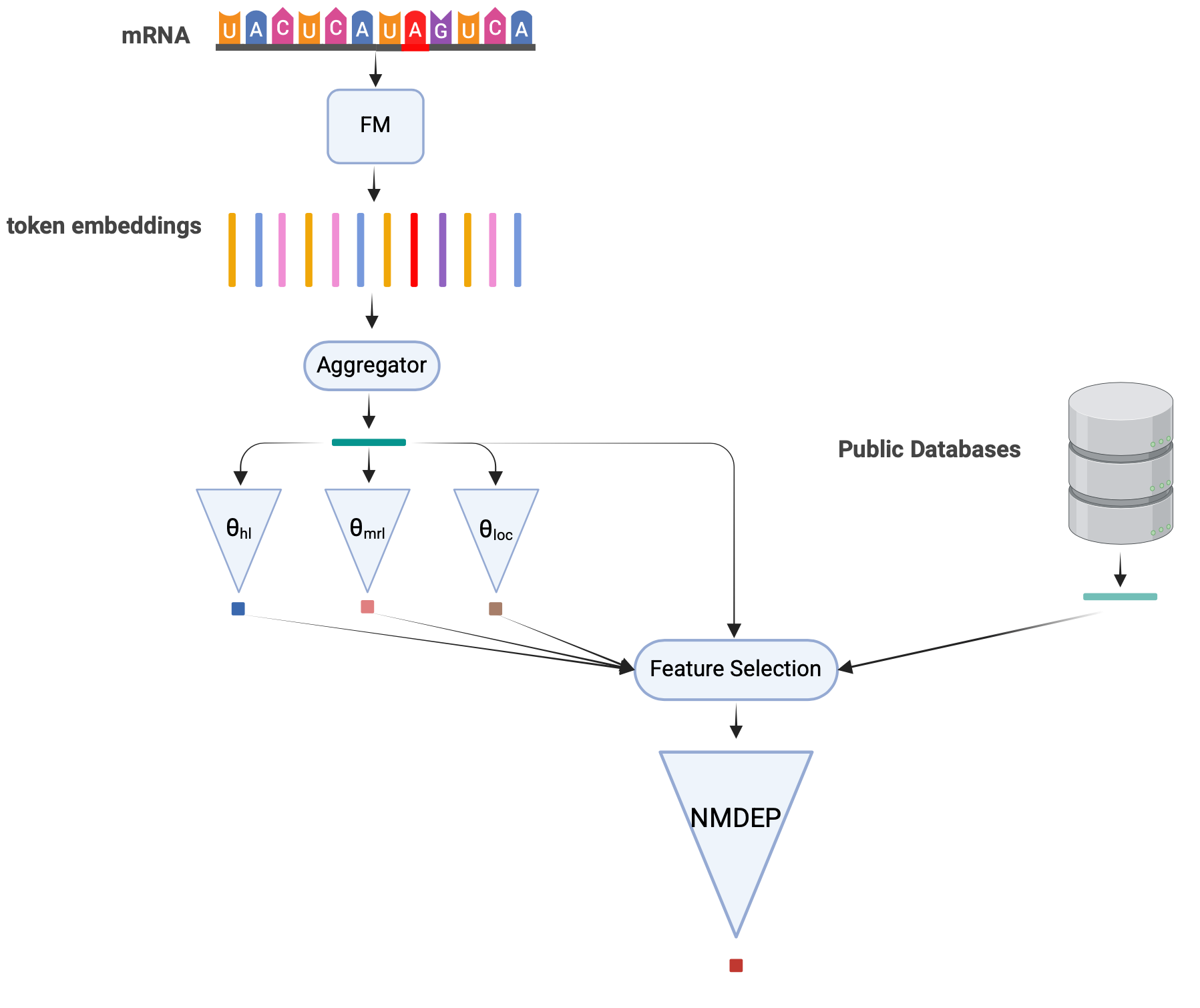} 
    \caption{\footnotesize NMDEP development overview: mRNA sequences are processed through a foundation model (FM) to extract token embeddings, which are averaged and used to predict half-life, ribosome loading, and localization. The selected features are then fed into NMDEP, a two-layer MLP with a hidden size of 8 neurons. Figure created with \url{BioRender.com}}
    \label{NMD_my_model}
\end{figure}

We developed NMDEP by optimizing rule-based thresholds, curating potentially relevant features, and imputing features with high missingness, including half-life, mean ribosome-leading, and RNA location. Figure \ref{NMD_my_model} provides an overview of the NMDEP workflow.

\paragraph{Feature threshold optimization}
\label{feature_optimize}

Three of the four baseline features (penultimate rule, close to start, and long exon) rely on predefined thresholds. To determine the optimal values, we conducted a two-step grid search, selecting thresholds that minimized validation loss. In the first step, we explored a broad range of options. After identifying the values with the lowest validation loss, we refined the search by performing a second, more focused grid search around the selected thresholds. The final parameter ranges and step sizes are summarized in Table \ref{grid_search_table}.

\begin{table}[h]
    \centering
    \begin{tabular}{lcc}
        \hline
        Feature & First Grid Search & Second Grid Search\\
        \hline
        Penultimate exon & range = [25, 75], step = 5 & range = [40, 50], step = 1 \\
        Start codon proximity & range = [50, 250], step = 25 & range = [75, 125], step = 5 \\
        Exon length & range = [350, 450], step = 25 & range = [350, 400], step = 5 \\
        \hline
    \end{tabular}
    \caption{Grid search parameter ranges and step sizes for feature threshold optimization.}
    \label{grid_search_table}
\end{table}

\paragraph{Feature annotation and selection}

For annotation, we extracted 76 potentially relevant features from public datasets, covering various aspects, including the position of the stop-gain variant, transcript and exon characteristics, evolutionary conservation and tolerance to variation, variant pathogenicity scores, intrinsic sequence properties, functional protein features, gene expression and regulation, and subcellular localization. Additionally, we incorporated the averaged mRNA Orthrus embeddings. For feature selection, we excluded features with low variance (\(< 0.01\)) or high correlation (\(> 0.8\)).

\paragraph{Auxiliary models}

Due to high missingness in three features including half-life (hl), mean ribosome loading (mrl), and subcellular localization we trained three auxiliary models to predict them using averaged Orthrus embeddings. This design is based on our benchmarking results (Table \ref{benchmarking_table}) which showed AggFirst approach with mean aggregator can perform well and is computationally efficient. Each auxiliary model is either an XGBoost regressor (for hl and mrl) or classifier (for localization), trained and tuned on 80\% of the data using cross-validation and evaluated on the remaining 20\%.

\paragraph{Model interpretation}

To interpret the model’s predictions, we used SHapley Additive exPlanations (SHAP) to assess global feature importance \citep{NIPS2017_7062}. SHAP quantifies each feature’s contribution by computing its average marginal impact across all possible feature combinations.

\paragraph{Genome-wide stop-gain variant simulation and NMDEP inference}

We used the Ensembl \citet{ensembl_simulate} to generate all possible single-nucleotide variants within coding regions. We then annotated these variants using VEP \citep{McLaren2016-ih} and retained only stop-gain variants in canonical transcripts. Finally, we applied NMDEP to estimate NMD efficiency for all retained variants.

\section{Results}
\label{results}

\subsection{Benchmarking sequence embedding representations}

Table \ref{benchmarking_table} summarizes the performance of the baseline model alongside three embedding-only approaches for NMD efficiency prediction. We observe that:  
1) None of the embedding-only models outperform the baseline.  
2) The choice of model, aggregation method, and sequence representation significantly affects performance.  
3) Among the non-baseline approaches, using the mean aggregator for ALT sequence embeddings yields the best improvement.  
4) DeepSet and AggLast offer performance gains but are computationally more demanding. 5) AggFirst, when combined with a mean aggregator and Alt sequence embeddings, achieves comparable performance while remaining computationally efficient.

\begin{table}[h]
    \centering
    \small
    \begin{tabular}{lllccccc}
        \toprule
        Model & Aggregator & Sequence & MAE $\downarrow$ & RMSE $\downarrow$ & $R^2$ $\uparrow$ & Spear. Corr $\uparrow$ & Pear. Corr $\uparrow$ \\
        \midrule
        \textbf{Baseline (4 rules)} & & & \textbf{0.8} & \textbf{1.07} & \textbf{0.35} & \textbf{0.67} & \textbf{0.6} \\
        \midrule
        \textbf{AggFirst} & mean & Alt & 0.9 & 1.17 & 0.15 & 0.45 & 0.41 \\
        & max & Alt & 1.09 & 1.33 & 0 & NA & NA \\
        & token & Alt & 0.99 & 1.27 & 0.01 & 0.27 & 0.2 \\
        & sum & Alt & 1.09 & 1.33 & 0 & NA & NA \\
        & mean & Alt - Ref & 0.94 & 1.22 & 0.08 & 0.39 & 0.32 \\
        & max & Alt - Ref & 1.06 & 1.31 & 0.03 & 0.24 & 0.2 \\
        & token & Alt - Ref & 1.07 & 1.31 & 0.02 & 0.2 & 0.15 \\
        & sum & Alt - Ref & 0.95 & 1.23 & 0.14 & 0.45 & 0.38 \\
        \midrule
        \textbf{AggLast} & mean & Alt & 0.89 & 1.17 & 0.16 & 0.47 & 0.44 \\
        & max & Alt & 0.94 & 1.28 & 0.06 & 0.46 & 0.45 \\
        & token & Alt & 1.04 & 1.29 & 0.06 & 0.31 & 0.24 \\
        & sum & Alt & 6.6 & 10.03 & -60.95 & 0.08 & 0.04 \\
        & mean & Alt - Ref & 0.93 & 1.18 & 0.14 & 0.42 & 0.39 \\
        & max & Alt - Ref & 1.04 & 1.31 & 0.02 & 0.19 & 0.21 \\
        & token & Alt - Ref & 1.03 & 1.30 & 0.04 & 0.24 & 0.21 \\
        & sum & Alt - Ref & 2.83 & 3.4 & -6.12 & 0.22 & 0.19 \\
        \midrule
        \textbf{DeepSet} & mean & Alt & 0.89 & 1.16 & 0.18 & 0.47 & 0.45 \\
        & max & Alt & 1 & 1.23 & 0.12 & 0.42 & 0.43 \\
        & token & Alt & 1.06 & 1.31 & 0.02 & 0.19 & 0.13 \\
        & sum & Alt & 1.1 & 1.45 & -0.29 & NA & NA \\
        & mean & Alt - Ref & 0.92 & 1.17 & 0.15 & 0.48 & 0.40 \\
        & max & Alt - Ref & 1 & 1.3 & 0 & 0.19 & 0.14 \\
        & token & Alt - Ref & 1.05 & 1.31 & 0.03 & 0.21 & 0.17 \\
        & sum & Alt - Ref & 1.04 & 1.3 & -0.04 & 0 & -0.05 \\
        \bottomrule
    \end{tabular}
    \caption{\footnotesize Benchmarking of embedding-only methods for NMD efficiency prediction. NA: not available.}
    \label{benchmarking_table}
\end{table}

\subsection{NMDEP training, evaluation, and application}

\paragraph{Feature optimization and selection}

We optimized the thresholds for three baseline features using a two-step grid search, both minimizing validation loss. The first search identified optimal thresholds over a broad range, while the second refined them around the best-performing values. The final thresholds are presented in Table \ref{grid_search_results}, with Figures \ref{grid_search_bigstep} and \ref{grid_search_smallstep} depicting the search process.

\begin{table}[h]
    \centering
    \begin{tabular}{lcc}
        \hline
        Feature& First Grid Search& Second Grid Search \\
        \hline
        Penultimate exon threshold & 45.0 & 49.0 \\
        Start codon proximity threshold & 100.0 & 120.0 \\
        Exon length threshold & 375.0 & 355.0 \\
        \hline
    \end{tabular}
    \caption{\footnotesize Optimized thresholds determined through a two-step grid search.}
    \label{grid_search_results}
\end{table}

Afterwards, we used these optimized features along with other potentially relevant feature (Table \ref{all_features}) and removed features with low variance or high correlation. This way we selected 60 features, as described in Table \ref{selected_features}.

\paragraph{Auxiliary models performances}

The performance of the auxiliary models on the test set varied across tasks. For half-life  prediction, the model achieved an MSE of 0.53, an MAE of 0.57, and a Spearman correlation of 0.68. For mean ribosome loading prediction, the model obtained an MSE of 0.61, an MAE of 0.63, and a Spearman correlation of 0.51. Finally, for subcellular localization classification, the model reached the accuracy of 0.70 and F1-score of 0.59.

\paragraph{NMDEP outperforms baseline and embedding-only models}

As shown in Table \ref{NMDEP_table}, NMDEP outperforms all other models, improving MAE by 10.7\%, RMSE by 9.2\%, \( R^2 \) by 24.4\%, Spearman correlation by 7\%, and Pearson correlation by 11\%.

\begin{table}[h]
    \centering
    \small
    \begin{tabular}{lccccc}
        \toprule
        Model & MAE $\downarrow$ & RMSE $\downarrow$ & $R^2$ $\uparrow$ & Spear. Corr $\uparrow$ & Pear. Corr $\uparrow$ \\
        \midrule
        \textbf{Baseline (4 rules)} & 0.8 & 1.07 & 0.35 & 0.67 & 0.6 \\
        \midrule
        \textbf{4 rules optimized} & 0.75 & 0.98 & 0.41 & 0.71 & 0.66 \\
        \textbf{Best of embedding-only models} & 0.89 & 1.16 & 0.18 & 0.48 & 0.45 \\
        \textbf{Features from \citet{Kim2024-eh}} & 0.78 & 1.06 & 0.3 & 0.63 & 0.56 \\
        \textbf{NMDEP} & \textbf{0.67} & \textbf{0.89} & \textbf{0.51} & \textbf{0.76} & \textbf{0.73} \\
        \bottomrule
    \end{tabular}
    \caption{\footnotesize Performance comparison of models for NMD efficiency prediction.}
    \label{NMDEP_table}
\end{table}

\paragraph{NMDEP feature interpretation}

\begin{figure}[h]
    \centering
    \includegraphics[width=0.85\textwidth]{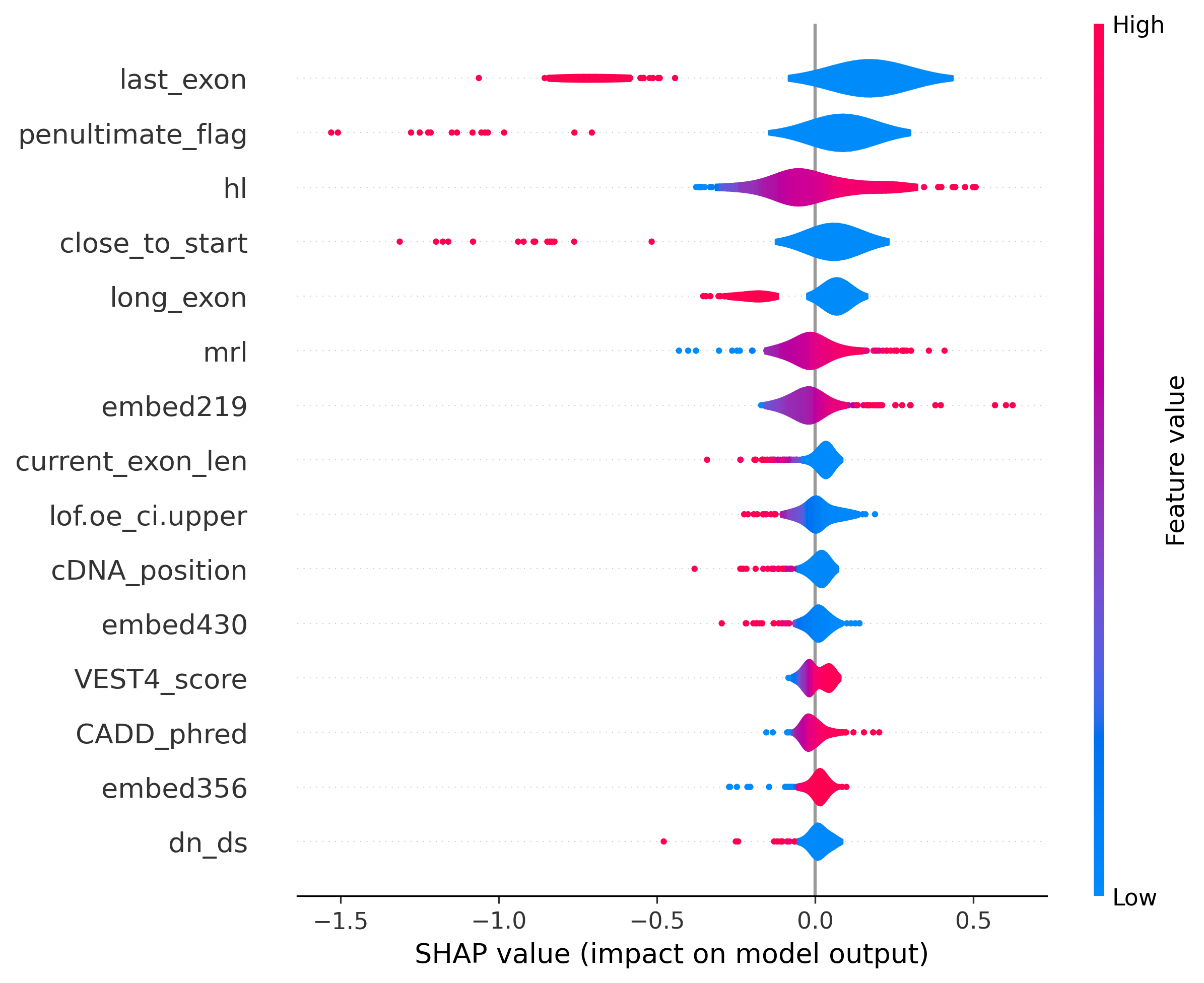} 
    \caption{\footnotesize Top 15 important features for NMDEP, ranked by SHAP values, which indicate each feature's impact on the model output.}
    \label{shap_violoin}
\end{figure}

\begin{figure}[h]
    \centering
    \includegraphics[width=0.85\textwidth]{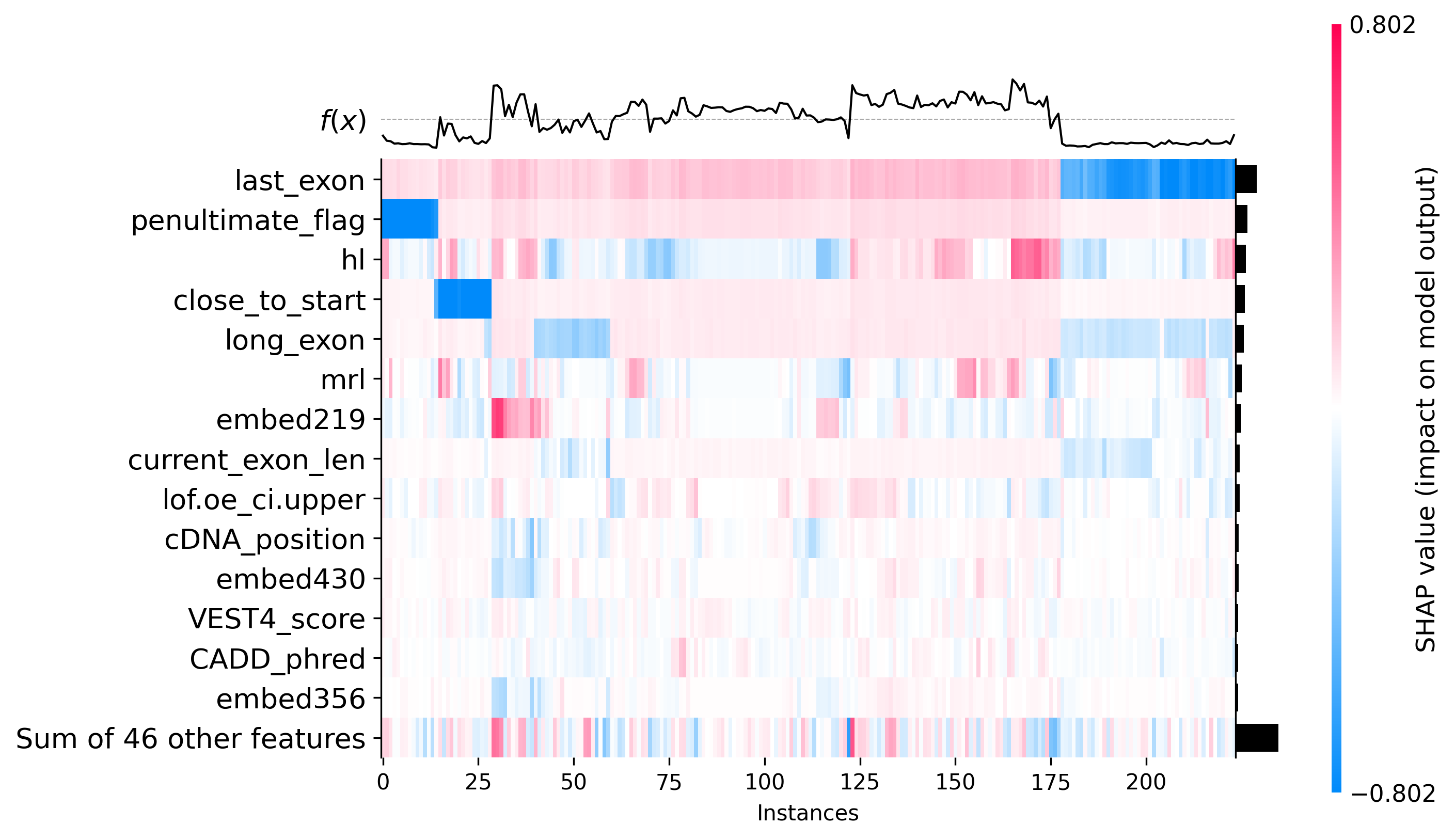} 
    \caption{\footnotesize Heatmap of SHAP values for all test set samples. \( f(x) \) is the model's output. Dark bars on the right indicate the overall contribution of each feature.}
    \label{shap_heatmap}
\end{figure}

Figure \ref{shap_violoin} presents the top 15 features, demonstrating that NMDEP effectively leverages previously known factors (e.g., the last exon and penultimate rule) while also identifying potentially interesting features (e.g., mean ribosome loading). Additionally, Figure \ref{shap_heatmap} visualizes the SHAP values of all features across instances in the test set, providing evidence that each feature contributes meaningfully to the model’s predictions.

\paragraph{Genome-wide NMDEP inference}

We simulated 2,921,293 stop-gain variants across 18,372 unique transcripts and applied NMDEP to estimate their NMD efficiency. To demonstrate the utility of NMDEP, we present predictions for two genes.

First, \textit{POLR3B}, a gene with many coding exons, exhibits variability in NMD efficiency across its sequence. Figure \ref{POLR3B_NMD} illustrates these variations, while Figure \ref{POLR3B_shap} provides a heatmap for interpretation, highlighting that proximity to the start codon and in-silico pathogenicity score (CADD\(_{\text{phred}}\)) are the most influential factors.

Second, \textit{TLR7}, which contains only a single coding exon, has consistently low NMD efficiency estimations, indicating that all stop-gain variants in this transcript can escape NMD (Figure \ref{TLR7_NMD}). The SHAP value heatmap (Figure \ref{TLR7_shap}) reveals that conservation scores are the primary determinant of low NMD efficiency across \textit{TLR7}.

\section{Discussion}

We developed NMDEP, a machine learning framework that combines sequence embeddings with curated biological features to predict NMD efficiency, outperforming rule-based and embedding-only models. Our benchmarking analysis demonstrated that models relying solely on sequence embeddings underperformed compared to simple rule-based heuristics, underscoring the need for biologically informed feature integration. NMDEP overcomes the limitations of existing approaches and provides a more accurate, generalizable solution for NMD efficiency prediction.

Our analysis confirmed known determinants of NMD, such as variant position, while also identifying novel contributors, such as ribosome loading. Additionally, NMDEP’s ability to predict NMD efficiency at single-variant resolution enabled a large-scale evaluation of more than 2.9 million simulated stop-gain variants, offering a systematic framework for assessing the impact of premature stop codons on transcript stability.

Despite its strong performance, NMDEP does not account for tissue-specific variations, which may limit its applicability across diverse biological contexts \citep{Palou-Marquez2024-jt}. Additionally, NMDEP may not fully capture the diverse regulatory mechanisms governing NMD.

Future work will enhance NMDEP by integrating tissue-specific NMD data for more precise predictions and incorporating frameshift and splicing effects to improve transcript stability modeling \citep{Litchfield2020-mv,McGlincy2008-tk}. Additionally, exploring its clinical utility in prioritizing pathogenic stop-gain variants will help translate predictions into real-world applications, advancing transcriptome regulation and disease understanding.


\bibliography{iclr2025_ai4na}
\bibliographystyle{iclr2025_conference}

\newpage

\appendix
\section{Appendix}

\newcounter{suppfigure}
\renewcommand{\thefigure}{S\arabic{suppfigure}} 
\setcounter{suppfigure}{1}
\renewcommand{\figurename}{Supplementary Figure}

\subsection{Supplementary figures}

\begin{figure}[h]
    \centering
    \includegraphics[width=0.95\textwidth]{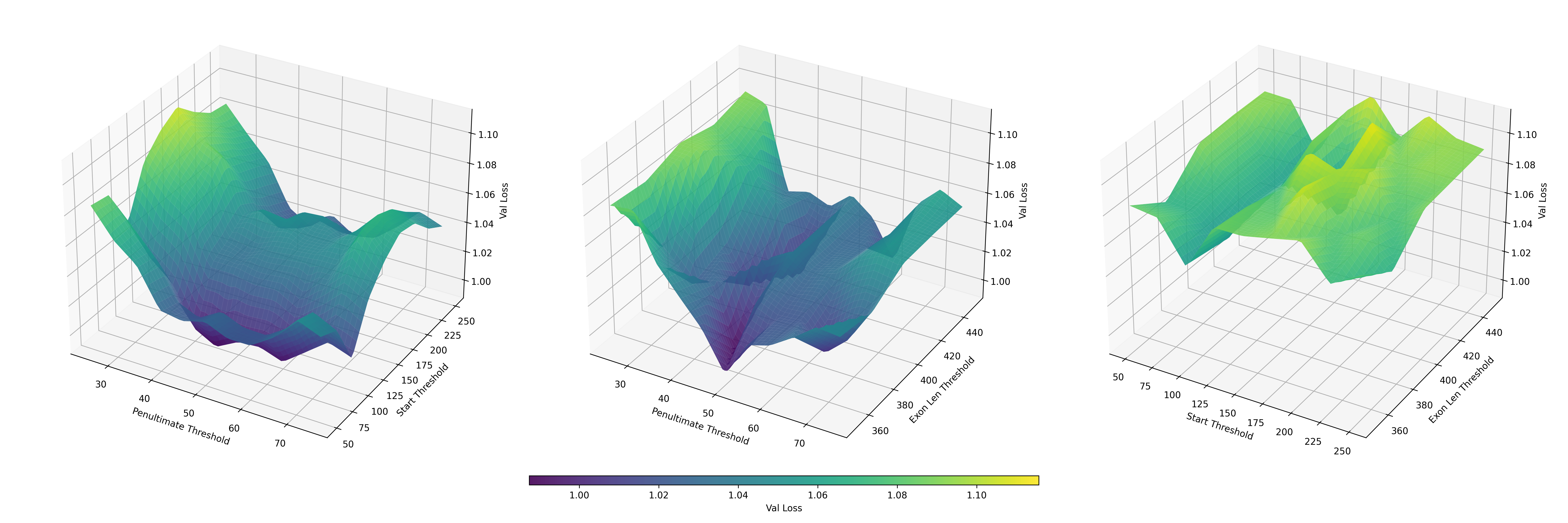} 
    \caption{\footnotesize First grid search results: Optimization of validation loss by varying thresholds.}
    \label{grid_search_bigstep}
    \refstepcounter{suppfigure}
\end{figure}

\begin{figure}[h]
    \centering
    \includegraphics[width=0.95\textwidth]{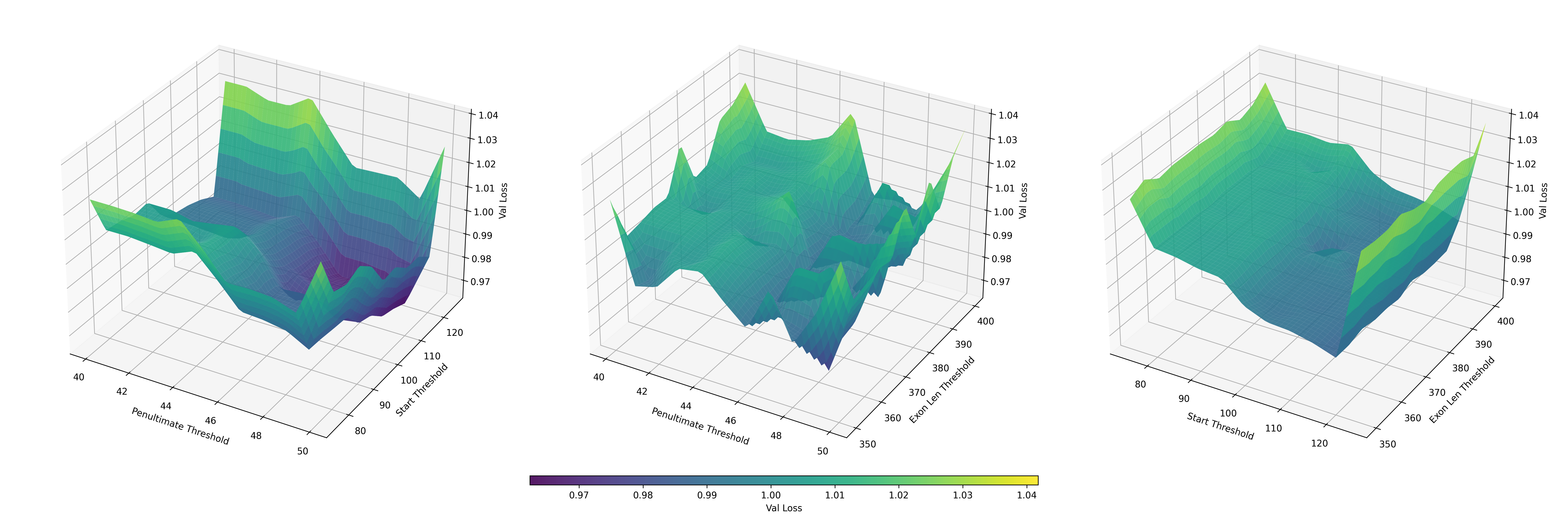} 
    \caption{\footnotesize Second grid search results: Refinement of validation loss optimization by adjusting thresholds based on the first grid search results.}
    \label{grid_search_smallstep}
    \refstepcounter{suppfigure}
\end{figure}

\newpage

\begin{figure}[h]
    \centering
    \includegraphics[width=0.95\textwidth]{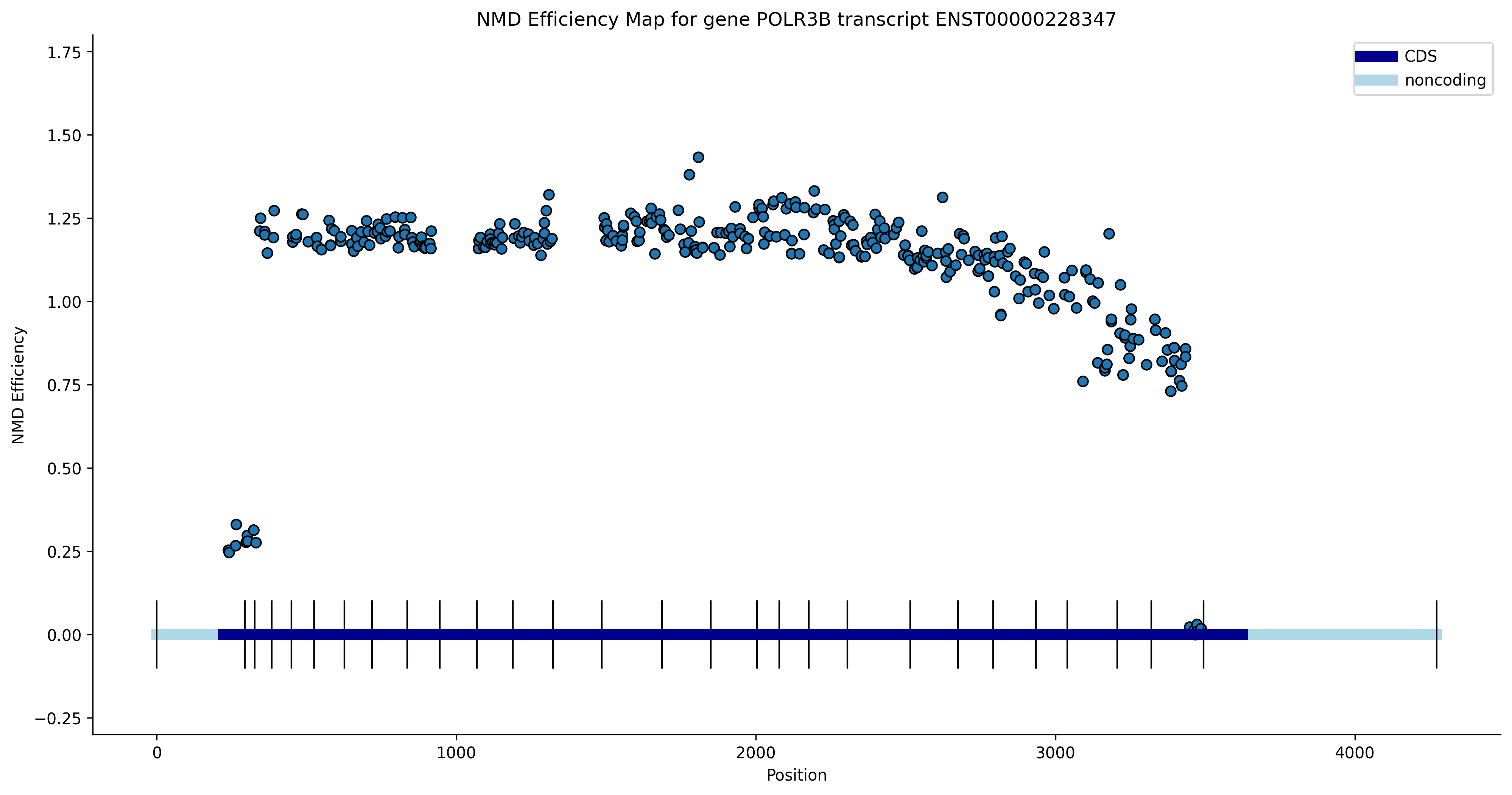} 
    \caption{\footnotesize NMDEP estimations for all simulated stop-gain variants across \textit{POLR3B}. Vertical lines indicate exon boundaries.}
    \label{POLR3B_NMD}
    \refstepcounter{suppfigure}
\end{figure}

\begin{figure}[h]
    \centering
    \includegraphics[width=0.95\textwidth]{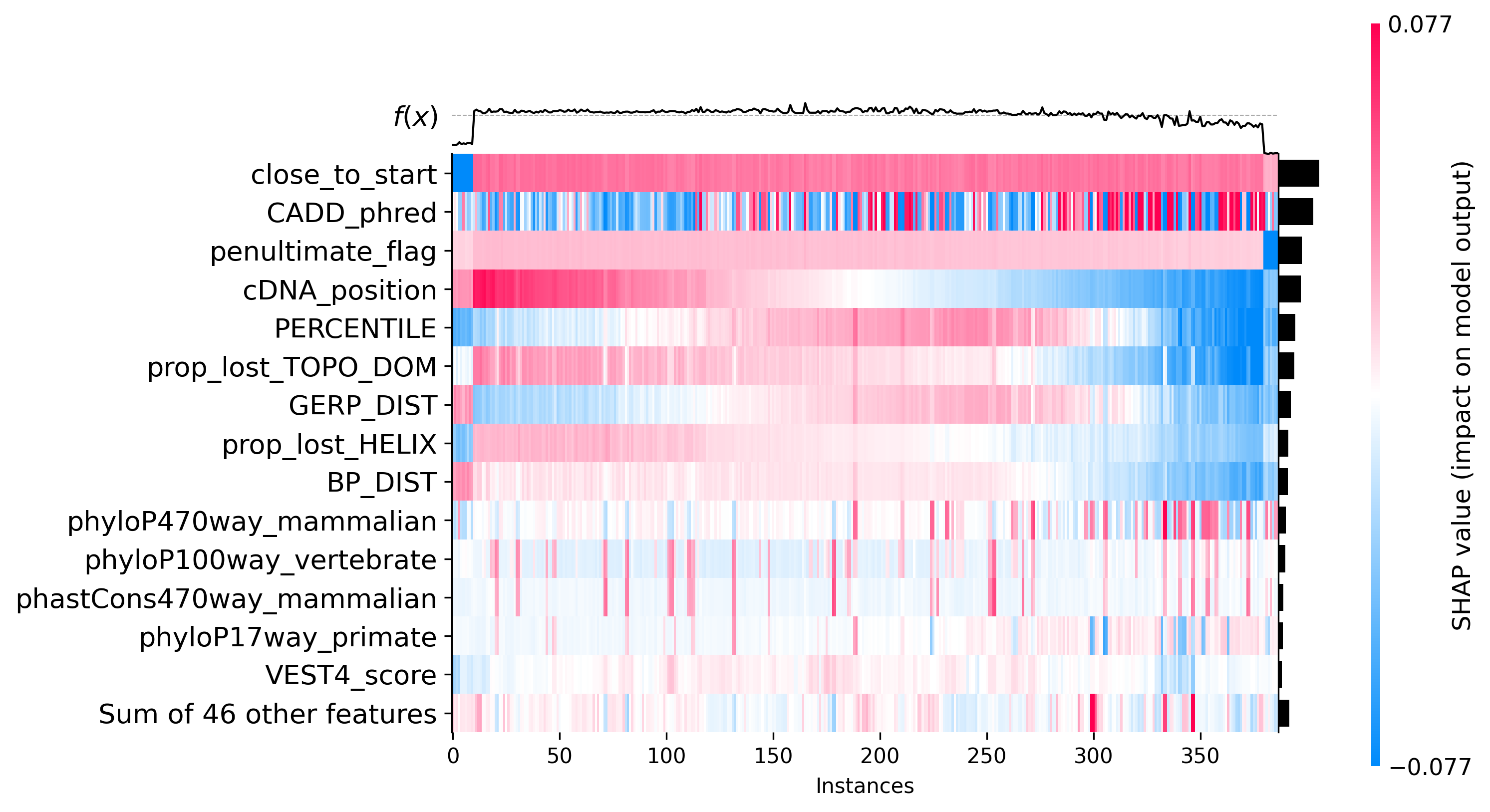} 
    \caption{\footnotesize Heatmap of SHAP values for all simulated stop-gain variants across \textit{POLR3B}. \( f(x) \) is the model's output. Dark bars on the right indicate the overall contribution of each feature.}
    \label{POLR3B_shap}
    \refstepcounter{suppfigure}
\end{figure}

\begin{figure}[h]
    \centering
    \includegraphics[width=0.95\textwidth]{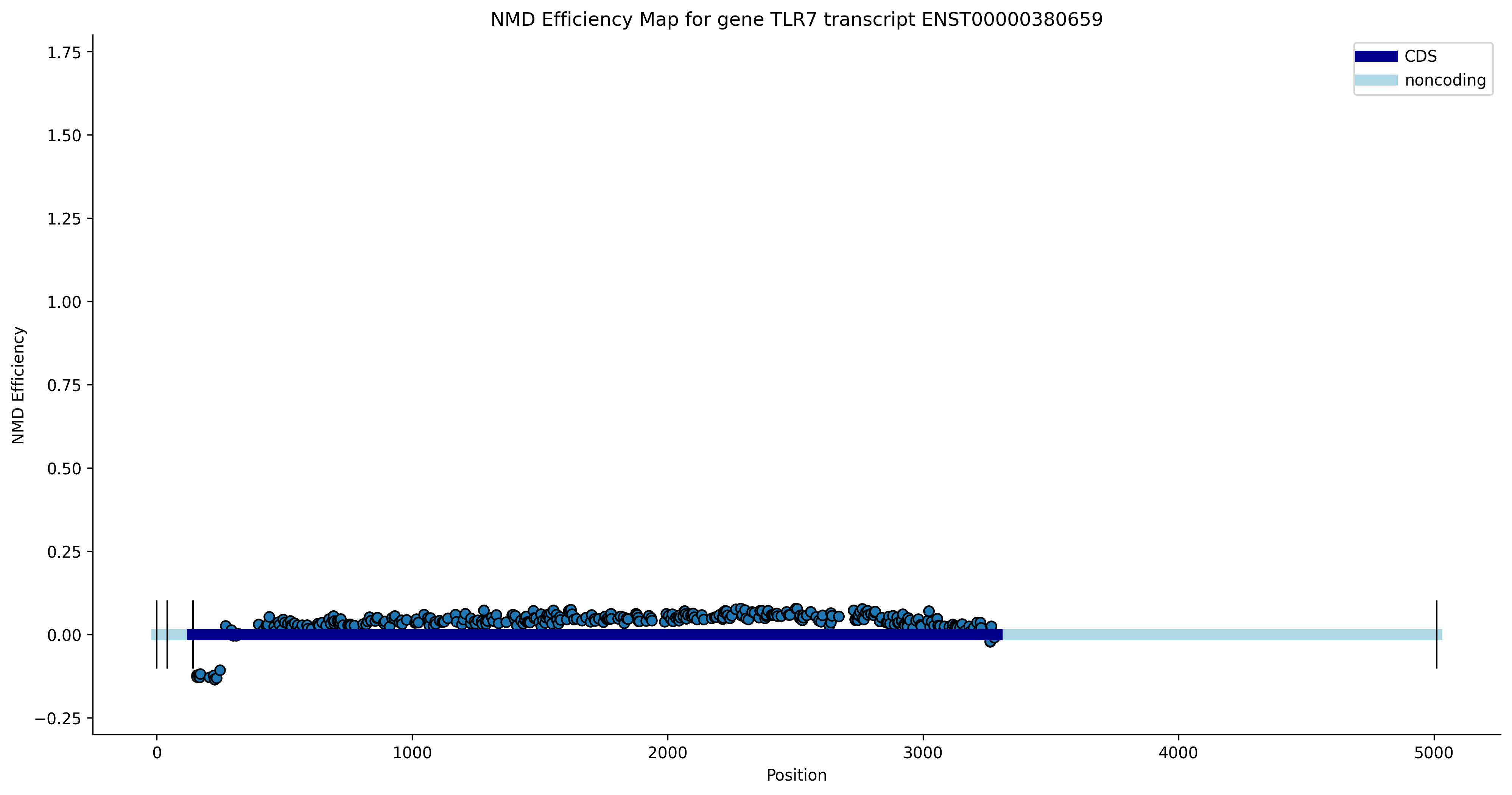} 
    \caption{\footnotesize NMDEP estimations for all simulated stop-gain variants across \textit{TLR7}. Vertical lines indicate exon boundaries.}
    \label{TLR7_NMD}
    \refstepcounter{suppfigure}
\end{figure}

\begin{figure}[h]
    \centering
    \includegraphics[width=0.95\textwidth]{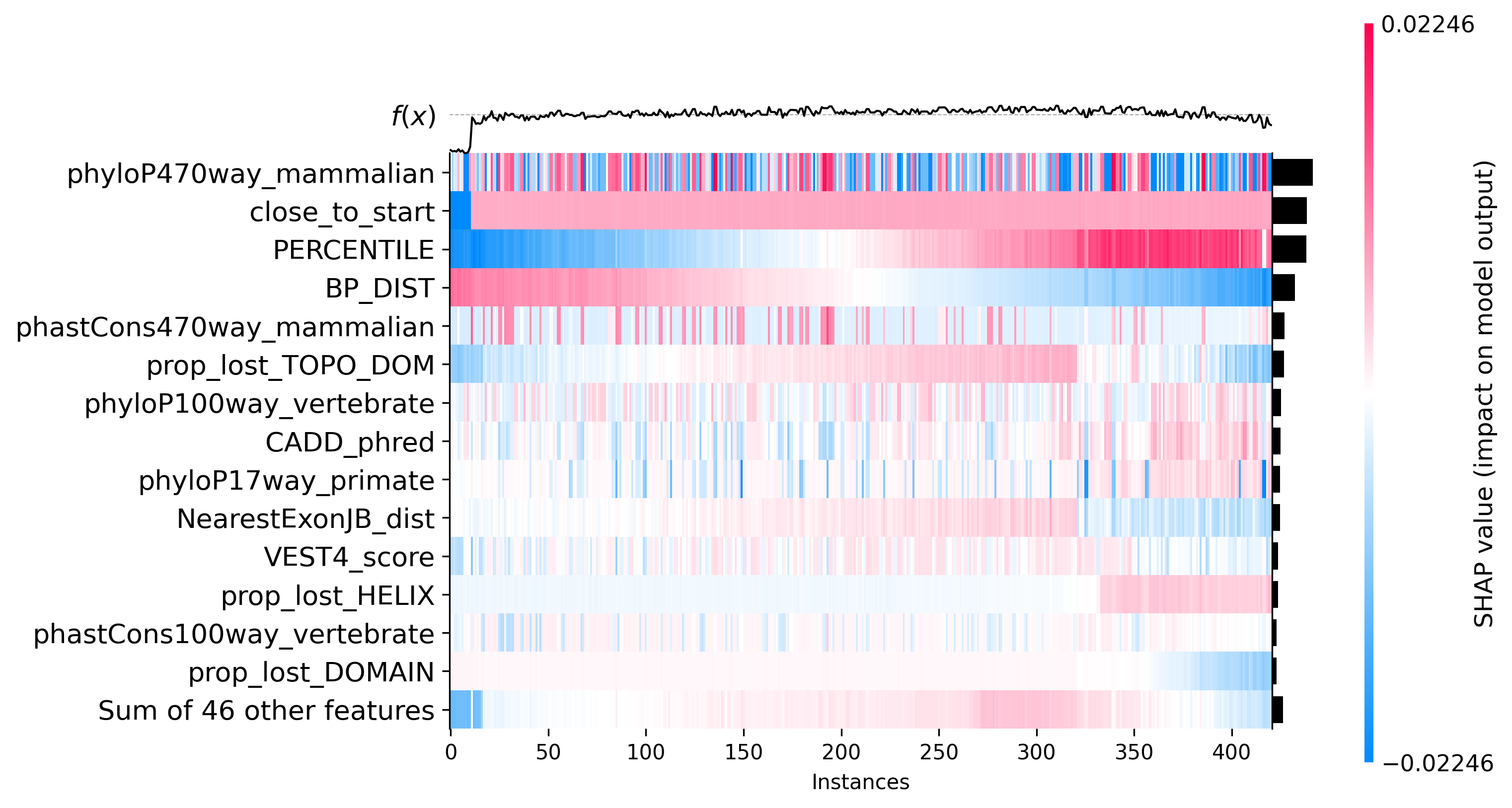} 
    \caption{\footnotesize Heatmap of SHAP values for all simulated stop-gain variants across \textit{TLR7}. \( f(x) \) is the model's output. Dark bars on the right indicate the overall contribution of each feature.}
    \label{TLR7_shap}
    \refstepcounter{suppfigure}
\end{figure}

\clearpage


\newcounter{supptable}
\renewcommand{\thetable}{S\arabic{supptable}} 
\setcounter{supptable}{1}
\renewcommand{\tablename}{Supplementary Table}

\subsection{Supplementary tables}

\begin{longtable}{|l|p{10cm}|}
    \hline
    \textbf{Feature}  & \textbf{Description} \\
    \hline
    \endfirsthead

    \hline
    \textbf{Feature}  & \textbf{Description} \\
    \hline
    \endhead

    \hline
    \endfoot

    \hline
    \endlastfoot

    cDNA\_position \citep{McLaren2016-ih}& Position of the mutation within the cDNA sequence. \\
CDS\_position \citep{McLaren2016-ih}&  Position of the mutation within the coding sequence. \\
Protein\_position \citep{McLaren2016-ih}&  Position of the mutation in the translated protein. \\
PERCENTILE \citep{Karczewski2020-lq}&  Percentile of the mutation’s position. \\
BP\_DIST \citep{Karczewski2020-lq}& Nucleotide distance from the variant position to the stop codon\\
GERP\_DIST \citep{Karczewski2020-lq}&  GERP-weighted distance from a variant position. \\
last\_exon &  Indicates if the mutation occurs in the last exon. \\
long\_exon &  Whether the affected exon is considered long. \\
penultimate\_flag &  Whether the mutation is in the last 49bp of second-to-last exon. \\
single\_exon &  Whether the gene has a single exon. \\
current\_exon\_number \citep{McLaren2016-ih}&  Exon number where the mutation is located. \\
total\_exon\_numbers \citep{McLaren2016-ih}&  Total number of exons in the gene. \\
DIST\_FROM\_LAST\_EXON \citep{Karczewski2020-lq}&  Distance of the mutation from the last exon. \\
NearestExonJB\_dist \citep{McLaren2016-ih}&  Distance to the nearest exon junction boundary. \\
close\_to\_start &  Whether the mutation is near the start codon. \\
utr5\_len \citep{genomekit}&  Length of the 5' UTR. \\
utr3\_len \citep{genomekit}&  Length of the 3' UTR. \\
total\_cds\_len \citep{genomekit}&  Total length of the coding sequence. \\
total\_exons\_len \citep{genomekit}&  Total length of all exons in the gene. \\
current\_exon\_len \citep{genomekit}&  Length of the exon where the mutation occurs. \\
hl \citep{Agarwal2022-ab}&  Estimated half-life of the mRNA. \\
mrl \citep{Sugimoto2022-da}&  mean ribosome load. \\
phyloP100way\_vertebrate \citep{Pollard2010-qu}&  Phylogenetic conservation across 100 vertebrate species. \\
phyloP17way\_primate \citep{Pollard2010-qu}&  Phylogenetic conservation across 17 primates. \\
phyloP470way\_mammalian \citep{Pollard2010-qu}&  Phylogenetic conservation across 470 mammals. \\
phastCons100way\_vertebrate \citep{Pollard2010-qu}&  Conservation score across 100 vertebrates. \\
phastCons17way\_primate \citep{Pollard2010-qu}&  Conservation score across 17 primates. \\
phastCons470way\_mammalian \citep{Pollard2010-qu}&  Conservation score across 470 mammals. \\
dn\_ds \citep{Jeffares2015-kx}&  Ratio of nonsynonymous to synonymous mutations. \\
lof.pLI \citep{Karczewski2020-lq}&  Probability of being loss-of-function intolerant. \\
lof.pRec \citep{Karczewski2020-lq}& Probability of intolerance to homozygous mutations. \\
lof.pNull \citep{Karczewski2020-lq}&   Probability of being tolerant to LoF variants. \\
mis.z\_score \citep{Karczewski2020-lq}&  Z-score for missense constraint. \\
syn.z\_score \citep{Karczewski2020-lq}&  Z-score for synonymous constraint. \\
lof.oe\_ci.upper\citep{Karczewski2020-lq} &  Upper bound of observed/expected LoF variation confidence interval. \\
VEST4\_score \citep{Carter2013-wh}& VEST4 pathogenicity predictor. \\
CADD\_phred \citep{Rentzsch2019-zz}&  CADD pathogenicity predictor. \\
fathmm-XF\_coding\_score \citep{Rogers2018-xz}& fathmm-XF pathogenicity predictor. \\
LoF\_HC \citep{Karczewski2020-lq}&  High-confidence loss-of-function (based on LOFTEE). \\
gnomad41\_genome\_AF \citep{Karczewski2020-lq}&  Allele frequency in the gnomAD genome dataset. \\
gnomad41\_exome\_AF \citep{Karczewski2020-lq}& Allele frequency in the gnomAD exome dataset. \\
CDS\_GC \citep{genomekit}&  GC content of the coding sequence. \\
UTR3\_GC \citep{genomekit}&  GC content of the 3' UTR. \\
UTR5\_GC \citep{genomekit}&  GC content of the 5' UTR. \\
embedX \citep{Fradkin2024-rr}&  Sequence embedding generated using Orthrus. X is between 0 to 511 \\
prop\_lost\_ACT\_SITE \citep{Saadat2024-tn}&  Proportion of active site lost due to mutation. \\
prop\_lost\_BINDING \citep{Saadat2024-tn}&  Proportion of binding sites lost. \\
prop\_lost\_COILED \citep{Saadat2024-tn}&  Proportion of coiled-coil regions lost. \\
prop\_lost\_COMPBIAS \citep{Saadat2024-tn}&  Proportion of compositionally biased regions lost. \\
prop\_lost\_DISULFID \citep{Saadat2024-tn}&  Proportion of disulfide bonds lost. \\
prop\_lost\_DOMAIN \citep{Saadat2024-tn}& Proportion of functional domains lost. \\
prop\_lost\_HELIX \citep{Saadat2024-tn}&  Proportion of alpha-helices lost. \\
prop\_lost\_MOD\_RES \citep{Saadat2024-tn}&  Proportion of modified residues lost. \\
prop\_lost\_MOTIF \citep{Saadat2024-tn}& Proportion of sequence motifs lost. \\
prop\_lost\_PROPEP \citep{Saadat2024-tn}& Proportion of propeptide regions lost. \\
prop\_lost\_REGION \citep{Saadat2024-tn}&  Proportion of functionally relevant regions lost. \\
prop\_lost\_REPEAT \citep{Saadat2024-tn}& Proportion of repeat regions lost. \\
prop\_lost\_SIGNAL \citep{Saadat2024-tn}&  Proportion of signal peptides lost. \\
prop\_lost\_STRAND \citep{Saadat2024-tn}&  Proportion of beta-strands lost. \\
prop\_lost\_TOPO\_DOM \citep{Saadat2024-tn}& Proportion of topological domains lost. \\
prop\_lost\_TRANSIT \citep{Saadat2024-tn}& Proportion of transit peptides lost. \\
prop\_lost\_TRANSMEM \citep{Saadat2024-tn}& Proportion of transmembrane regions lost. \\
prop\_lost\_TURN \citep{Saadat2024-tn}&  Proportion of beta-turns lost. \\
prop\_lost\_ZN\_FING \citep{Saadat2024-tn}& Proportion of zinc finger domains lost. \\
abundance \citep{Zeng2024-xb}&  abundance level. \\
exp\_var \citep{Zeng2024-xb}&  Variability of RNA expression across tissues. \\
tau \citep{Zeng2024-xb}& Tissue specificity of gene expression. \\
connectedness \citep{Zeng2024-xb}&  Connectivity of the gene in coexpression networks. \\
betweenness \citep{Zeng2024-xb}& Centrality of the gene in protein-protein interaction networks. \\
TF \citep{Zeng2024-xb}&  Whether the gene encodes a transcription factor. \\
Nucleus \citep{Cui2022-bn}&  localization in the nucleus. \\
Exosome \citep{Cui2022-bn}& localization in exosomes. \\
Cytosol \citep{Cui2022-bn}& localizationin the cytosol. \\
Cytoplasm \citep{Cui2022-bn}& localization in the cytoplasm. \\
Ribosome \citep{Cui2022-bn}& localization in ribosomes. \\
Membrane \citep{Cui2022-bn}&  localization is membrane-bound. \\
Endoplasmic\_reticulum \citep{Cui2022-bn}& localization in the endoplasmic reticulum. \\
\caption{\footnotesize List of initially annotated features.}
\label{all_features} 
\refstepcounter{supptable}
\end{longtable}

\begin{longtable}{|l|l|p{10cm}|}    
    \hline
    \textbf{Feature} & \textbf{Category} & \textbf{Description} \\
    \hline
    \endfirsthead

    \hline
    \textbf{Feature} & \textbf{Category} & \textbf{Description} \\
    \hline
    \endhead

    \hline
    \endfoot

    \hline
    \endlastfoot

    last\_exon & Transcript and Exon Features & Indicates if the mutation occurs in the last exon. \\
    long\_exon & Transcript and Exon Features & Whether the affected exon is considered long. \\
    penultimate\_flag & Transcript and Exon Features & Whether the mutation is in the last 49bp of second-to-last exon. \\
    close\_to\_start & Transcript and Exon Features & Whether the mutation is near the start codon. \\
    hl & Transcript and Exon Features & Estimated half-life of the mRNA. \\
    mrl & Transcript and Exon Features & Mean ribosome load. \\
    cDNA\_position & Position of Stop-Gain Variant & Position of the mutation within the cDNA sequence. \\
    PERCENTILE & Position of Stop-Gain Variant & Percentile of the mutation’s position. \\
    GERP\_DIST & Position of Stop-Gain Variant & GERP-weighted distance from a variant position. \\
    BP\_DIST & Position of Stop-Gain Variant & Nucleotide distance from the variant position to the stop codon. \\
    utr5\_len & Transcript and Exon Features & Length of the 5' UTR. \\
    current\_exon\_len & Transcript and Exon Features & Length of the exon where the mutation occurs. \\
    VEST4\_score & Variant Pathogenicity & VEST4 pathogenicity predictor. \\
    CADD\_phred & Variant Pathogenicity & CADD pathogenicity predictor. \\
    phyloP100way\_vertebrate & Conservation and Tolerance & Phylogenetic conservation across 100 vertebrate species. \\
    dn\_ds & Conservation and Tolerance & Ratio of nonsynonymous to synonymous mutations. \\
    abundance & Expression and Regulation & Abundance level. \\
    shet & Conservation and Tolerance & Posterior mean for shet. \\
    lof.oe\_ci.upper & Conservation and Tolerance & Upper bound of observed/expected LoF variation confidence interval. \\
    lof.pRec & Conservation and Tolerance & Probability of intolerance to homozygous mutations. \\
    UTR5\_GC & Intrinsic Sequence Features & GC content of the 5' UTR. \\
    connectedness & Expression and Regulation & Connectivity of the gene in coexpression networks. \\
    prop\_lost\_DOMAIN & Functional Protein Features & Proportion of functional domains lost. \\
    prop\_lost\_HELIX & Functional Protein Features & Proportion of alpha-helices lost. \\
    prop\_lost\_REGION & Functional Protein Features & Proportion of functionally relevant regions lost. \\
    prop\_lost\_TOPO\_DOM & Functional Protein Features & Proportion of topological domains lost. \\
    LoF\_HC & Variant Pathogenicity & High-confidence loss-of-function (based on LOFTEE). \\
    NearestExonJB\_dist & Transcript and Exon Features & Distance to the nearest exon junction boundary. \\
    TF & Expression and Regulation & Whether the gene encodes a transcription factor. \\
    tau & Expression and Regulation & Tissue specificity of gene expression. \\
    phyloP17way\_primate & Conservation and Tolerance & Phylogenetic conservation across 17 primates. \\
    phyloP470way\_mammalian & Conservation and Tolerance & Phylogenetic conservation across 470 mammals. \\
    phastCons100way\_vertebrate & Conservation and Tolerance & Conservation score across 100 vertebrates. \\
    phastCons17way\_primate & Conservation and Tolerance & Conservation score across 17 primates. \\
    phastCons470way\_mammalian & Conservation and Tolerance & Conservation score across 470 mammals. \\
    mis.z\_score & Conservation and Tolerance & Z-score for missense constraint. \\
    syn.z\_score & Conservation and Tolerance & Z-score for synonymous constraint. \\
    lof.pNull & Conservation and Tolerance & Probability of being tolerant to LoF variants. \\
    exp\_var & Expression and Regulation & Variability of RNA expression across tissues. \\
    utr3\_len & Transcript and Exon Features & Length of the 3' UTR. \\
    total\_exons\_len & Transcript and Exon Features & Total length of all exons in the gene. \\
    fathmm-XF\_coding\_score & Variant Pathogenicity & fathmm-XF pathogenicity predictor. \\
    Nucleus & Subcellular Localization & Localization in the nucleus. \\
    Cytosol & Subcellular Localization & Localization in the cytosol. \\
    Cytoplasm & Subcellular Localization & Localization in the cytoplasm. \\
    Ribosome & Subcellular Localization & Localization in ribosomes. \\
    Membrane & Subcellular Localization & Localization is membrane-bound. \\
    Endoplasmic\_reticulum & Subcellular Localization & Localization in the endoplasmic reticulum. \\
    embed6 & Intrinsic Sequence Features & Sequence embedding 6 from Orthrus. \\
    embed8 & Intrinsic Sequence Features & Sequence embedding 8 from Orthrus. \\
    embed90 & Intrinsic Sequence Features & Sequence embedding 90 from Orthrus. \\
    embed145 & Intrinsic Sequence Features & Sequence embedding 145 from Orthrus. \\
    embed205 & Intrinsic Sequence Features & Sequence embedding 205 from Orthrus. \\
    embed219 & Intrinsic Sequence Features & Sequence embedding 219 from Orthrus. \\
    embed230 & Intrinsic Sequence Features & Sequence embedding 230 from Orthrus. \\
    embed240 & Intrinsic Sequence Features & Sequence embedding 240 from Orthrus. \\
    embed254 & Intrinsic Sequence Features & Sequence embedding 254 from Orthrus. \\
    embed309 & Intrinsic Sequence Features & Sequence embedding 309 from Orthrus. \\
    embed356 & Intrinsic Sequence Features & Sequence embedding 356 from Orthrus. \\
    embed430 & Intrinsic Sequence Features & Sequence embedding 430 from Orthrus. \\
    \caption{\footnotesize List of selected features and their descriptions.}
    \label{selected_features}
    
    \refstepcounter{supptable}
    
\end{longtable}

\end{document}